\documentclass[RNAAS]{aastex62}

\begin{document}

\title{Discovery of four young asteroid families}

\correspondingauthor{Bojan Novakovi\' c}
\email{bojan@matf.bg.ac.rs}

\author[0000-0001-6349-6881]{Bojan Novakovi\' c}
\affiliation{Department of Astronomy, Faculty of Mathematics, University of Belgrade \\
Studentski trg 16, 11000 Belgrade, Serbia \\}

\author{Viktor Radovi\' c}
\affiliation{Department of Astronomy, Faculty of Mathematics, University of Belgrade \\
Studentski trg 16, 11000 Belgrade, Serbia \\}



\section{Introduction}

Asteroid families are groups of objects that share similar orbital parameters, and in most cases,
also similar spectral characteristics \citep{Milani2014Icar,nes2015}. These groups are thought to originate 
from a single parent asteroid, and are closly related to many aspects of asteroid-based 
research \citep{cellino2009,Hsieh2018AJ}.

Once created, families however undergo dynamical and collisional evolution \citep{nov2015ApJ,bottke2015}, that both start immediately after the brake-up event. For this reason, a specially interesting 
subgroup of asteroid families are young ones, i.e. those formed less than 10 Myrs ago. They
have attracted a lot of attention since the first such family has been discovered by \citet{nes2002}.
About 20 young families have been discovered so far, but new discoveries are still very important. 
Here we report discovery of four new young asteroid families.
 
\section{Methods}

Asteroid families are usually identified in the space of proper orbital elements \citep{KM2003}, 
using the Hierarchical Clustering Method (HCM) proposed by \citet{Zappala1990}. For identification
of young families, a somewhat different approach is more suitable. 
Backward integration method (BIM) is a standard tool to estimate ages of young asteroid families \citep{nes2003}, that could be also used to verify a common origin of asteroid groups \citep{nov2012MNRAS_mbc}. 

The BIM is based on the fact that immediately after the formation,
members of an asteroid family have nearly the same orbital elements. These elements are then
dispersed over time, and signatures of their clustering are erased. However, for young families, 
the epoch of past convergence of secular angles may still be revealed by propagated their orbits
backward in time \citep{Radovic2017}.

Recently we have produced the largest catalog\footnote{The catalog is available at \href{http://asteroids.matf.bg.ac.rs/fam/properelements.php}{Asteroid Families Portal} \citep{Viktor2017MNRAS}} of asteroid proper orbital elements so far, that provides these data for more than 630,000 objects. As the procedure to compute proper elements includes numerical integrations of asteroids' orbits, these are also available for the same objects.

Taking advantage of the availability of these data-sets, our strategy to identify young families consists of two steps. In the first step, we applied the HCM to the new catalog of asteroid proper elements to identify all small groups recognizable at velocity cut-off distance of 15 m/s. In the second step, we used the BIM to confirm the common origin and the young age of newly identified clusters.

\section{Results}

Using the above described methodology we have identified four new young dynamical families.
Three of them are located in the inner asteroid belt, namely the (525)~Adelaide,
(6142)~Tantawi and (18429)~1994AO1, while the family of (2258)~Viipuri is
located in the middle part of the asteroid belt.

\begin{figure}[ht!]
\begin{center}
\includegraphics[scale=0.6,angle=-90]{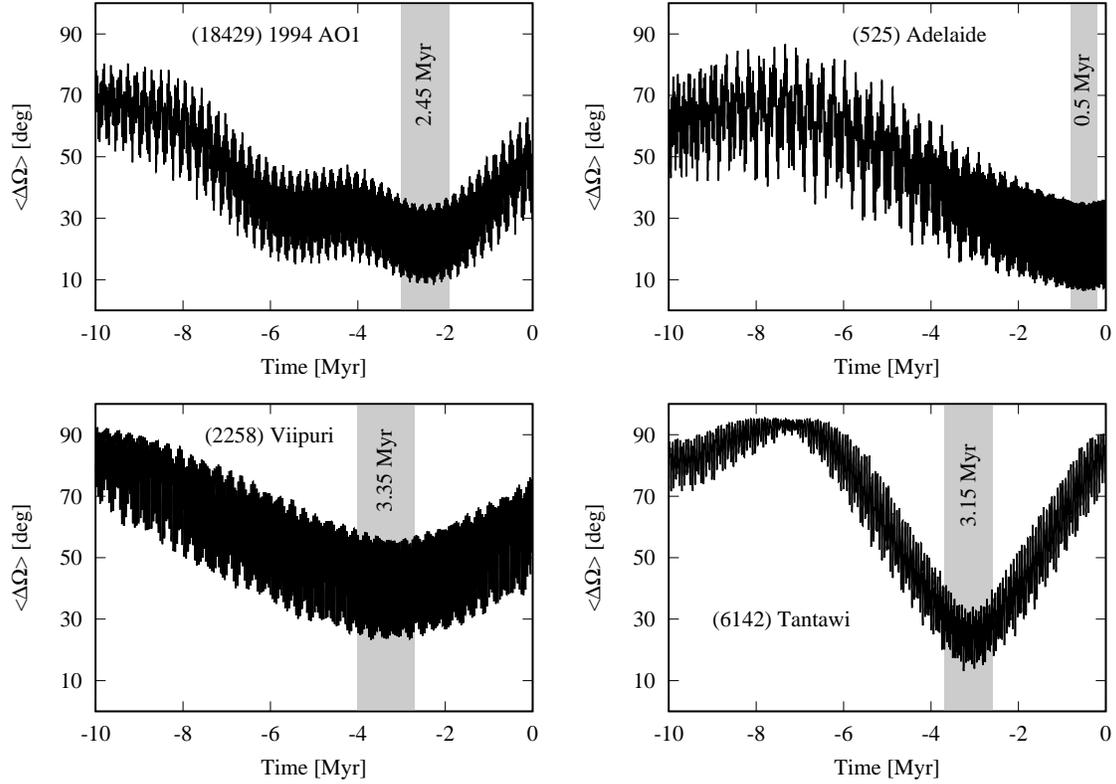}
\caption{The time-evolution of the average differences in the mean nodal longitudes for members of four new young asteroid families. The names of the families and their approximate ages are indicated in each of the panels.} \label{fig:1}
\end{center}
\end{figure}

The differences in secular orbital angles of potential members of the new families
are shown in Fig.~\ref{fig:1}. In each case, a tight clustering of nodal longitudes is
clearly visible at some point. This suggests that asteroids from each of the groups 
have a common origin, and allows a preliminary estimation of their ages (Fig.~\ref{fig:1}). 

Three new clusters consist of dark primitive objects, except for the Adelaide cluster whose 
members are asteroids probably of taxonomic S-type.
Two of these clusters are subfamilies of larger and older families.
Based on the data from \citet{nes2015PDSS}, the Adelaide cluster is located within the Flora family, 
while the Tantawi group is a sub-family of the Nysa-Polana complex.

\section{Discussion} 

The clustering of orbital angles that we have found for each of the new families suggests that their members
originate from the common parent bodies. However, the exact mechanism of their formation is 
still to be determined. While large older families are formed by collisions, in the recent years it was shown that small young families may also be formed by rotational fission of critically spinning parent bodies \citep{pravec2018}.

The ages of the new families obtained here are determined using the nominal orbits of their members only,
and purely gravitational model. These results should be improved by including the Yarkovsky effect in the dynamical model, and taking into account the orbital uncertainties \citep{nov2012lorre}.

\acknowledgments

This work has been supported by the Ministry of Education, Science and Technological Development of the Republic of Serbia, under the Projects 176011 and III44006.

\end{document}